\documentclass[twoside,twocolumn,9pt]{article}
\pdfoutput=1
\usepackage{extsizes}
\usepackage[super,sort&compress,comma]{natbib} 
\usepackage[version=3]{mhchem}
\usepackage[left=1.5cm, right=1.5cm, top=1.785cm, bottom=2.0cm]{geometry}
\usepackage{balance}
\usepackage{times,mathptmx}
\usepackage{sectsty}
\usepackage{notes2bib}
\usepackage{graphicx}
\usepackage{lastpage}
\usepackage[format=plain,justification=justified,singlelinecheck=false,font={stretch=1.125,small,sf},labelfont=bf,labelsep=space]{caption}
\usepackage{float}
\usepackage{fancyhdr}
\usepackage{fnpos}
\usepackage[english]{babel}
\addto{\captionsenglish}{%
	\renewcommand{\refname}{Notes and references}
}
\usepackage{array}
\usepackage{droidsans}
\usepackage{charter}
\usepackage[T1]{fontenc}
\usepackage[usenames,dvipsnames]{xcolor}
\usepackage{setspace}
\usepackage[compact]{titlesec}
\usepackage{hyperref}

\usepackage{siunitx}

\usepackage{epstopdf}

\definecolor{cream}{RGB}{222,217,201}

\begin{document}
	
	\pagestyle{fancy}
	\thispagestyle{plain}
	\fancypagestyle{plain}{
		
		\fancyhead[C]{\includegraphics[width=18.5cm]{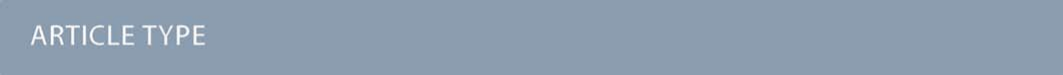}}
		\fancyhead[L]{\hspace{0cm}\vspace{1.5cm}\includegraphics[height=30pt]{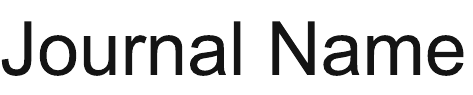}}
		\fancyhead[R]{\hspace{0cm}\vspace{1.7cm}\includegraphics[height=55pt]{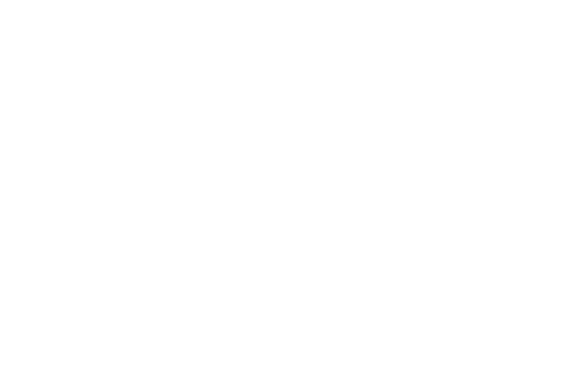}}
		\renewcommand{\headrulewidth}{0pt}
	}
	
	\makeFNbottom
	\makeatletter
	\renewcommand\LARGE{\@setfontsize\LARGE{15pt}{17}}
	\renewcommand\Large{\@setfontsize\Large{12pt}{14}}
	\renewcommand\large{\@setfontsize\large{10pt}{12}}
	\renewcommand\footnotesize{\@setfontsize\footnotesize{7pt}{10}}
	\makeatother
	
	\renewcommand{\thefootnote}{\fnsymbol{footnote}}
	\renewcommand\footnoterule{\vspace*{1pt}%
		\color{cream}\hrule width 3.5in height 0.4pt \color{black}\vspace*{5pt}} 
	\setcounter{secnumdepth}{5}
	
	\makeatletter 
	\renewcommand\@biblabel[1]{#1}            
	\renewcommand\@makefntext[1]%
	{\noindent\makebox[0pt][r]{\@thefnmark\,}#1}
	\makeatother 
	\renewcommand{\figurename}{\small{Fig.}~}
	\sectionfont{\sffamily\Large}
	\subsectionfont{\normalsize}
	\subsubsectionfont{\bf}
	\setstretch{1.125} 
	\setlength{\skip\footins}{0.8cm}
	\setlength{\footnotesep}{0.25cm}
	\setlength{\jot}{10pt}
	\titlespacing*{\section}{0pt}{4pt}{4pt}
	\titlespacing*{\subsection}{0pt}{15pt}{1pt}
	
	\fancyfoot{}
	\fancyfoot[LO,RE]{\vspace{-7.1pt}\includegraphics[height=9pt]{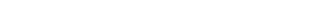}}
	\fancyfoot[CO]{\vspace{-7.1pt}\hspace{13.2cm}\includegraphics{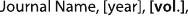}}
	\fancyfoot[CE]{\vspace{-7.2pt}\hspace{-14.2cm}\includegraphics{RF}}
	\fancyfoot[RO]{\footnotesize{\sffamily{1--\pageref{LastPage} ~\textbar  \hspace{2pt}\thepage}}}
	\fancyfoot[LE]{\footnotesize{\sffamily{\thepage~\textbar\hspace{3.45cm} 1--\pageref{LastPage}}}}
	\fancyhead{}
	\renewcommand{\headrulewidth}{0pt} 
	\renewcommand{\footrulewidth}{0pt}
	\setlength{\arrayrulewidth}{1pt}
	\setlength{\columnsep}{6.5mm}
	\setlength\bibsep{1pt}
	
	\graphicspath{{SpeedDisp1D_Draft/}}
	
	\makeatletter 
	\newlength{\figrulesep} 
	\setlength{\figrulesep}{0.5\textfloatsep} 
	
	\newcommand{\topfigrule}{\vspace*{-1pt}%
		\noindent{\color{cream}\rule[-\figrulesep]{\columnwidth}{1.5pt}} }
	
	\newcommand{\botfigrule}{\vspace*{-2pt}%
		\noindent{\color{cream}\rule[\figrulesep]{\columnwidth}{1.5pt}} }
	
	\newcommand{\dblfigrule}{\vspace*{-1pt}%
		\noindent{\color{cream}\rule[-\figrulesep]{\textwidth}{1.5pt}} }
	
	\makeatother
	
	\twocolumn[
	\begin{@twocolumnfalse}
		\vspace{3cm}
		\sffamily
		\begin{tabular}{m{4.5cm} p{13.5cm} }
			\includegraphics{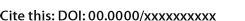} & \noindent\LARGE{\textbf{Diversity of self-propulsion speeds reduces motility-induced clustering in confined active matter
}} \\
			\vspace{0.3cm} & \vspace{0.3cm} \\
			
			& \noindent\large{Pablo de Castro$^{\ast a}$, Francisco M.\ Rocha$^{b}$, Saulo Diles$^{c}$, Rodrigo Soto$^{a}$ and Peter Sollich$^{d, e}$} \\
			
			\includegraphics{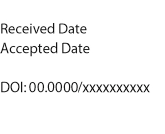} & \noindent\normalsize{
				
				Self-propelled swimmers such as bacteria agglomerate into clusters as a result of their persistent motion. In 1D, those clusters do not coalesce macroscopically and the stationary cluster size distribution (CSD) takes an exponential form. We develop a minimal lattice model for active particles in narrow channels to study how clustering is affected by the interplay between \textit{self-propulsion speed diversity} and confinement. A mixture of run-and-tumble particles with a distribution of self-propulsion speeds is simulated in 1D.  Particles can swap positions at rates proportional to their relative self-propulsion speed. Without swapping, we find that the average cluster size $L_\text{c}$ \textit{decreases} with diversity and follows a non-arithmetic power mean of the single-component $L_\text{c}$'s, unlike the case of tumbling-rate diversity previously studied. Effectively, the mixture is thus equivalent to a system of identical particles whose self-propulsion speed is the \textit{harmonic} mean self-propulsion speed of the mixture.  With swapping, particles escape more quickly from clusters. As a consequence, $L_\text{c}$ decreases with swapping rates and depends less strongly on diversity. We derive a dynamical equilibrium theory for the CSDs of binary and fully polydisperse systems. Similarly to the clustering behaviour of one-component models, our qualitative results for mixtures are expected to be universal across active matter. Using literature experimental values for the self-propulsion speed diversity of unicellular swimmers known as choanoflagellates, which naturally differentiate into slower and faster cells, we predict that the error in estimating their $L_\text{c}$ via one-component models which use the conventional arithmetic mean self-propulsion speed is around $30\%$.

			} \\
			
		\end{tabular}
		
	\end{@twocolumnfalse} \vspace{0.6cm}
	
	]
	
	\renewcommand*\rmdefault{bch}\normalfont\upshape
	\rmfamily
	\section*{}
	\vspace{-1cm}


	\footnotetext{\textit{$^{a}$Departamento de F\'{i}sica, Facultad de Ciencias F\'{i}sicas y Matem\'{a}ticas, Universidad de Chile, Avenida Blanco Encalada 2008, Santiago, Chile. Email: pdecastro@ing.uchile.cl}}
	\footnotetext{\textit{$^{b}$Aix Marseille University, CNRS, IUSTI, 13453 Marseille, France}}
	\footnotetext{\textit{$^{c}$Faculdade de F\'isica, Universidade Federal do Par\'a, Campus Salin\'opolis, Rua Raimundo Santana Cruz S/N, 68721-000, Salin\'opolis, Par\'a, Brazil}}
	\footnotetext{\textit{$^{d}$Disordered Systems Group, Department of Mathematics, King's College London, London, United Kingdom}}
	\footnotetext{\textit{$^{e}$Institut f\"ur Theoretische Physik, Georg-August-Universit\"at G\"ottingen, 37077 G\"ottingen, Germany}}
	
	

\section{Introduction}
A collection of self-propelled particles can spontaneously separate into dense and dilute regions even without attractive forces. This process, known as motility-induced phase separation (MIPS),\cite{cates2015motility,soto2014run,sepulveda2016coarsening,slowman2016jamming} occurs if the propulsion direction is sufficiently persistent against stochasticity, in which case there is enough time for the particles to trap each other and form large clusters.\cite{ginot2018aggregation,redner2016classical} For one-dimensional (1D) systems, such as fertilizing bacteria living in long narrow soil pores,\cite{quelas2016swimming, ranjard2001quantitative,mannik2009bacterial,fuentes2005bacterial} MIPS generates an exponentially decaying stationary cluster size distribution (CSD) as shown for run-and-tumble (RT), active Brownian, and active Ornstein-Uhlenbeck particles.\cite{D0SM00687D,dandekar2020hard,vanhille2019collective,caprini2020time,PhysRevE.90.062301,ao2014active,ghosh2014giant,ao2015diffusion,barberis2019phase} When the self-propellers are sufficiently strong to push each other, the clusters themselves can also move persistently,\cite{illien2020speed, barberis2019phase} producing deviations from a simple exponential CSD.\cite{vanhille2019collective}. In two-dimensional (2D) or quasi-two-dimensional systems, an exponential CSD modulated by a power law arises as observed in experiments with bacteria,\cite{peruani2012collective,zhang2010collective,be2020phase,keymer2008computation} in experiments and simulations with active colloids,\cite{buttinoni2013dynamical,levis2014clustering} and in lattice models.\cite{soto2014run} The CSDs can also be affected by solvent-mediated hydrodynamic interactions between the particles.\cite{alarcon2017morphology}

In a typical bacteria population there is a broad dispersion of motility parameters, i.e.\ the bacteria are not identical swimmers.\cite{andrea2020,berg2008coli,de2021active} However, for simplicity, effects of \textit{motility diversity} on CSDs are usually overlooked.\cite{bechinger2016active}
For RT bacteria, one can consider that the tumbling rate or the self-propulsion speed (or both) is not the same for all particles, that is, the system has a distribution of motility parameters. The fact that in a population of bacteria different self-propulsion speeds are found is a result of their different ages, reproduction stages, shapes, sizes, running modes, etc.\cite{ipina2019bacteria,berg2008coli,sparacino2020solitary,berdakin2013quantifying} For both passive and active fluids, `diversity' of some particle attribute can fundamentally change phase behaviour.\cite{PabloPeter1,warren1999phase,PabloPeter2,PabloPeter3,decastro2019,stenhammar2015activity,kolb2020active,hoell2019multi,wittkowski2017nonequilibrium,takatori2015theory,grosberg2015nonequilibrium,curatolo2020cooperative,wang2020phase,van2020predicting,dolai2018phase,lin2021phase}
In Ref.~\citenum{de2021active}, of which the present work can be regarded as a companion paper, we considered a multi-component mixture of RT particles on a 1D discrete lattice (as a simple model for active particles in narrow channels \cite{angelani2017confined,costanzo2012transport,costanzo2014motility,wu2018transport,ao2014active,wang2021preferred,caprini2021collective,daddi2018state,bisht2020rectification,mannik2010bacteria,mannik2009bacterial,wioland2016directed,locatelli2015active,ketzetzi2021activity,aguilar2018collective,gravish2015glass}) interacting only via excluded volume, i.e.\ they do not push each other. The distinct particle types were characterized by their own tumbling rates as inspired by experimental observations with \textit{Escherichia coli}; these experiments show that the tumbling rate of each bacterium changes stochastically but slowly, leading approximately to a log-normal distribution of constant tumbling rates for the system.\cite{figueroa20183d,andrea2020} Particles moving directly towards each other were allowed to cross at a constant rate, therefore mimicking the effects of a ``soft'' confinement where particles can swap their positions along the quasi-1D channel. Interactions via biochemical signalling were assumed negligible or absent. We observed an exponential CSD with an average cluster size $L_\text{c}$ that \textit{increases} with tumbling-rate diversity. This clustering amplification phenomenon is induced solely by tumbling-rate diversity as the global average tumbling rate remains fixed in the analysis. On the other hand, by relaxing the confinement, large cluster sizes are reduced and tumbling-rate diversity becomes less important. Furthermore, tumbling-rate diversity generates an average cluster size $L_\text{c}$ that is given by an arithmetic average of the $L_\text{c}$'s that the single-component systems would have at the same global density.

\begin{figure}[!h]
	\centering
	\includegraphics[width=0.95\columnwidth]{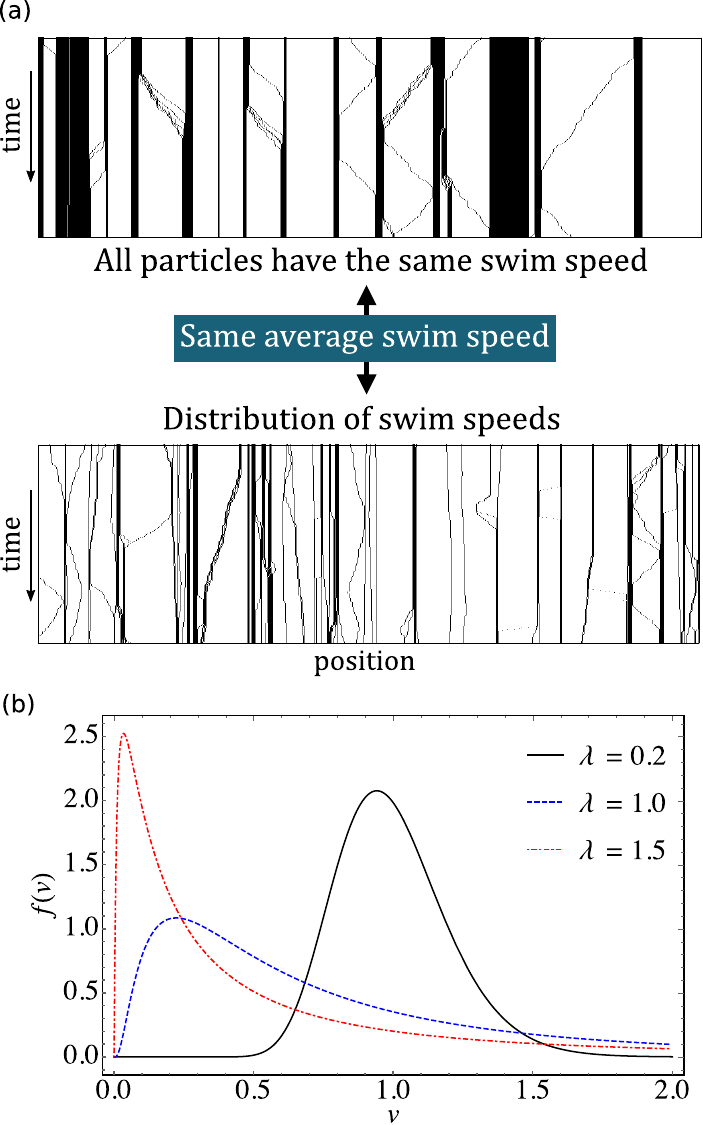}
	\caption{(a) Steady state of a monodisperse system (top) of run-and-tumble particles on a 1D lattice compared against its mixture counterpart with polydispersity in the self-propulsion speeds (bottom); particles cannot swap ($p_\text{s}=0$). Only $1000$ of $N=10^4$ simulated lattice sites are shown. Considering the entire simulated system, the global concentration and the average self-propulsion speed in both cases are the same: dimensionless concentration $\phi=0.2$,  tumbling rate $\alpha=0.005$, and average self-propulsion speed $\langle v\rangle=1$. In the mixture case (bottom) the polydisperse distribution parameter [see eqn~\eqref{polydist}] is $\lambda=2$. Time flows downwards along the vertical axis for $300$ time steps. Positions are on the horizontal axis. Vacancies are in white and particles in black. (b) Polydisperse distribution of self-propulsion speeds, eqn~\eqref{polydist}, for three different distribution parameters $\lambda$ at fixed average self-propulsion speed $\langle v\rangle=v_0\exp{(\lambda^2/2)}=1$.} 
	\label{Fig1}
\end{figure}

The self-propulsion speeds in Ref.~\citenum{de2021active} were set the same for all particles. The motility diversity was therefore entirely encoded into the tumbling-rate distribution. In the present work we examine the complementary and fundamentally distinct case of self-propulsion speed diversity and its effects on the stationary cluster size distributions of active particles in narrow environments: particles differ only in their self-propulsion speed but are assigned the same tumbling rate. To isolate the effects of self-propulsion speed diversity, we consider a distribution of self-propulsion speeds whose system average is kept fixed while only its variance is tuned. We find that the average cluster size \textit{decreases} with the variance of the self-propulsion speed distribution if one keeps everything else fixed, including the distribution average; see Fig.~\ref{Fig1}. Also, here we consider the clustering effects of particle ``overtaking'', that is, faster particles can overtake slower ones---if moving in the same direction---with a rate proportional to their relative self-propulsion speed. 

Since in the monodisperse case (i.e., without diversity) $L_\text{c}$ depends on the motility parameters self-propulsion speed and tumbling rate (denoted by $v$ and $\alpha$, respectively) as $L_\text{c}\sim\sqrt{v/\alpha}$, one could na\"ively think that knowing $L_\text{c}$ for tumbling-rate diversity automatically provides a prescription for obtaining $L_\text{c}$ in the case of self-propulsion speed diversity (hereafter referred to just as \textit{speed diversity}). We show here that this is not true. By employing an arithmetic average of the $L_\text{c}$'s of the single-component systems, one would still get the correct qualitative result that $L_\text{c}$ \textit{decreases} with speed diversity, but the simulation values for $L_\text{c}$ presented below are much lower than the arithmetic average, implying that this type of average is quantitatively inadequate to describe speed diversity.

We notice that Ref.~\citenum{marconi2016velocity} considers off-lattice active particles in 2D whose self-propulsion speeds vary both in time and between particles, following a Gaussian colored noise. In a certain way, this also corresponds to having diversity of self-propulsion speeds, but not ``quenched'' in time as we do below. Furthermore, the parameter used by the authors of Ref.~\citenum{marconi2016velocity} to increase the variance of the self-propulsion speed distribution also increases the average self-propulsion speed, i.e., these two distribution parameters cannot be decoupled in their analysis. Finally, in Ref.~\citenum{marconi2016velocity} the cluster sizes are not calculated. Similarly, we mention that, for off-lattice 2D binary mixtures of fast and slow active Brownian particles, motility-induced stationary average cluster sizes have been briefly discussed in Ref.~\citenum{kolb2020active}. In particular, the authors showed numerically that by reducing the \textit{ratio} between the slower and faster self-propulsion speeds, the average cluster sizes decrease. This could be regarded as a somewhat expected result since by reducing any active speed of the problem without increasing the rest of them, activity-induced clustering should indeed be reduced. This is crucially different from our work as here we decrease the slower speeds by the same amount that we increase the faster speeds. In our analysis we (i) consider RT particles on a lattice, (ii) analytically calculate cluster sizes, (iii) obtain new insights for confined systems, and (iv) probe quantitatively how clustering depends solely on self-propulsion speed diversity.

This paper is organized as follows. In Section~\ref{model} our lattice model is presented. Sections \ref{simulations} and \ref{theory} contain our main numerical and analytical results, respectively, for each model variant: (i) binary mixture, (ii) binary mixture with particle swapping, and (iii) fully polydisperse mixture. Section~\ref{conc} gives our conclusions and discussion.

\section{Model}
\label{model}
Let us review the RT model presented in Ref.~\citenum{soto2014run}, where all particles are identical. Consider a 1D discrete lattice with $N$ sites and periodic boundary conditions. The maximum occupancy per site is one. Each particle has a propulsion director, which can be left or right. The total number of particles is $M = \phi N$, where $\phi$ is the dimensionless global particle concentration. The initial positions are all distinct and otherwise random. Each particle is also given an initial random director. In each time step, $M$ individual particle updates are performed. The update algorithm is as follows. A particle is selected at random and a new director for this particle is chosen at random, with probability $\alpha$ (hereafter we refer to $\alpha$ either as a probability or as a ``rate'', and similarly for the probabilities of other elementary processes occurring at each time step). Thus, the probability to have a \emph{different} director is $\alpha/2$. A tumble event occurs if the particle changes its director. Otherwise, the particle preserves its previous director. Next, if the propulsion director points towards a neighbouring empty site, then the particle moves to this new position. A particle is then chosen anew. The updates are sequential. Our units are such that the lattice spacing and the time step are fixed to unity. A particle can be chosen more than once in a single time step, so that the swim speed (i.e.\ the free-space speed acquired by the self-propelled particle in the run mode) of the mobile particles fluctuates around unity. To consider a nominal swim speed $v\neq1$, all one needs to do is to perform $v M$ particle updates, that is, an average of $v$ updates per particle. In this case, we divide $\alpha$ by $v$ to prevent the tumbling rate from scaling with the self-propulsion speed. 
In this model, there is neither imposed Vicsek-like velocity alignment \cite{vicsek1995novel} nor the spontaneous velocity alignment shown to arise due to interparticle forces in off-lattice systems of active particles \cite{caprini2020spontaneous} in narrow circular channels, where, for high persistence, particles can start rotating collectively \cite{caprini2021collective}. Also, although active particles in a channel may push large clusters for long distances,\cite{debnath2020enhanced} it is not completely clear to which extent this is reflected in experimental CSDs.\cite{vanhille2019collective,peruani2012collective} Such effects are not considered here. Finally, a cluster is defined as a contiguous group of occupied sites.

The above model is monodisperse: all particles have the same motility properties, i.e.\ they have identical swim speed and tumbling rate. Here we consider a more realistic scenario where the particles have distinct swim speeds, while their tumbling rates continue to be monodisperse. To simulate a system of particles of different speeds, we use the following approach. A selected particle that has been assigned a fixed speed $v$ will be updated if $v$ is larger than a random variable between $0$ and $v_{\rm max}$, a simulation cutoff parameter. Otherwise, the selection is discarded. Once the total amount of individual particle updates is $\langle v \rangle M$, where $\langle v \rangle$ is the average particle speed in the sense of an arithmetic average, the time step ends. By using this procedure, the number of updates per particle per time step is $\langle v \rangle$, and faster (slower) particles are more (less) likely to be updated (and therefore to move), proportionally to their nominal speeds. Here, too, we divide $\alpha$ by the selected particle's $v$ so that the tumbling rate does not scale up with the self-propulsion speed. 
For a binary mixture, we consider half the particles with swim speed $v_A=v_0(1+\delta)$ and the other half with speed $v_B=v_0(1-\delta)$. As we vary the degree of speed diversity, $\delta$, the average speed $\langle v \rangle=v_0$ remains fixed. We use specifically $v_{\rm max}=2 v_0$, as that is the largest value the larger speed $v_A$ can have (for $\delta=1$), and set $v_0=1$. The speeds are assigned so that the initial state is randomly homogeneous and well-mixed. For simplicity, we do not present simulation data for mixing proportions other than $50$-$50\%$. This is because the continuous distribution case described below already covers a more general situation, although the binary theory in Section \ref{theory} was also validated via simulations with different global proportions. 

We also consider a fully polydisperse system, i.e.\ with a continuous distribution of speeds. We choose a log-normal distribution, which is adequate for non-negative variables. Also, it corresponds to the same shape as the distribution of tumbling rates in \textit{E.~coli} bacteria\cite{de2021active} and is visually similar (see Fig.~\ref{Fig1}b) to the experimental speed distributions of many swimming microorganisms, \textit{E.~coli} included.\cite{ipina2019bacteria,berg2008coli,sparacino2020solitary,berdakin2013quantifying} Our normalized fully-polydisperse distribution thus reads
\begin{equation}
	f(v)=\frac{1}{\sqrt{2 \pi }   \lambda v}\exp{\left(-\frac{\left[\log\left(v/v_0\right)\right]^2}{2 \lambda ^2}\right)},
	\label{polydist}
\end{equation}
where $v_0$ and $\lambda$ are the distribution parameters. We keep $\langle v \rangle\equiv\int_{0}^{\infty}v\,f(v)\,\textrm{d}v=v_0\exp{(\lambda^2/2)}$ fixed while the polydispersity degree is changed by varying $\lambda$. The monodisperse case corresponds to the limit $\lambda\to0$. For the simulations, the cutoff $v_{\rm max}=4\langle v \rangle$ was found to be sufficiently large. In any case, according to the theory in Section \ref{theory}, the particular functional form or parameters of the speed distribution do not affect our main qualitative results as described below.

To mimic the effects of a narrow channel whose width allows for neighbouring bacteria to swap positions, we proceed as follows. Consider a particle of type $i$ after it has potentially tumbled and moved but before a new particle selection occurs. 
At this stage of the algorithm, we allow for \textit{head-to-head crossing}, i.e.\ 
the particle will exchange positions with its neighbour with a probability $p_{\textrm{s}}$ if, and only if, their directors point towards each other. Here $p_{\textrm{s}}$ is a constant rate such that an increase in channel width corresponds to an increase in $p_{\textrm{s}}$. 
With this algorithm, out of all possible head-to-head crossing events involving particles of types $i$ and $j$ within a given time step, a fraction approximately equal to $p_{\textrm{s}}(v_i+v_j)$ will indeed occur, on average. The scaling with the speeds occurs because the number of particle updates is proportional to the speed. Here, we do not divide $p_s$ by speed: it is physically reasonable that the effective swapping rates are proportional to the relative speed since the self-propulsion forces of bacteria are typically proportional to their self-propulsion speeds. Similarly, we allow for \textit{overtaking}, i.e.\  particle $i$ will swap positions with its neighbouring particle $j$ with a probability proportional to $v_i-v_j$ if, and only if, their directors point in the same directions and $i$ is behind $j$ with $v_i>v_j$. Since in our algorithm the possibility of overtaking is considered only when the faster particle is selected, we set the nominal overtaking swapping rate to be $p_{\textrm{s}}(v_i-v_j)/v_i$: otherwise the effective swapping rate would scale with the square of the speeds. In summary, faster particles can overtake slower ones at an effective rate proportional to their relative speed. 

\section{Simulations}
\label{simulations}
Our stationary numerical results were obtained from simulations with periodic boundary conditions and $N=2000$ sites, except where a proper sampling of the swim speed distribution requires a larger system, in which case we used $N=10^4$. For visualization, we recorded snapshots of the steady-state system for $300$ successive time steps after $t=10^7$. The CSD and other similarly averaged quantities were calculated from $9000$ uncorrelated configurations recorded every $10^4$ time steps from $t=10^7$ onwards, within the same simulation.

\subsection{Binary mixture}
\begin{figure}[!h]
	\centering
	\includegraphics[width=0.95\columnwidth]{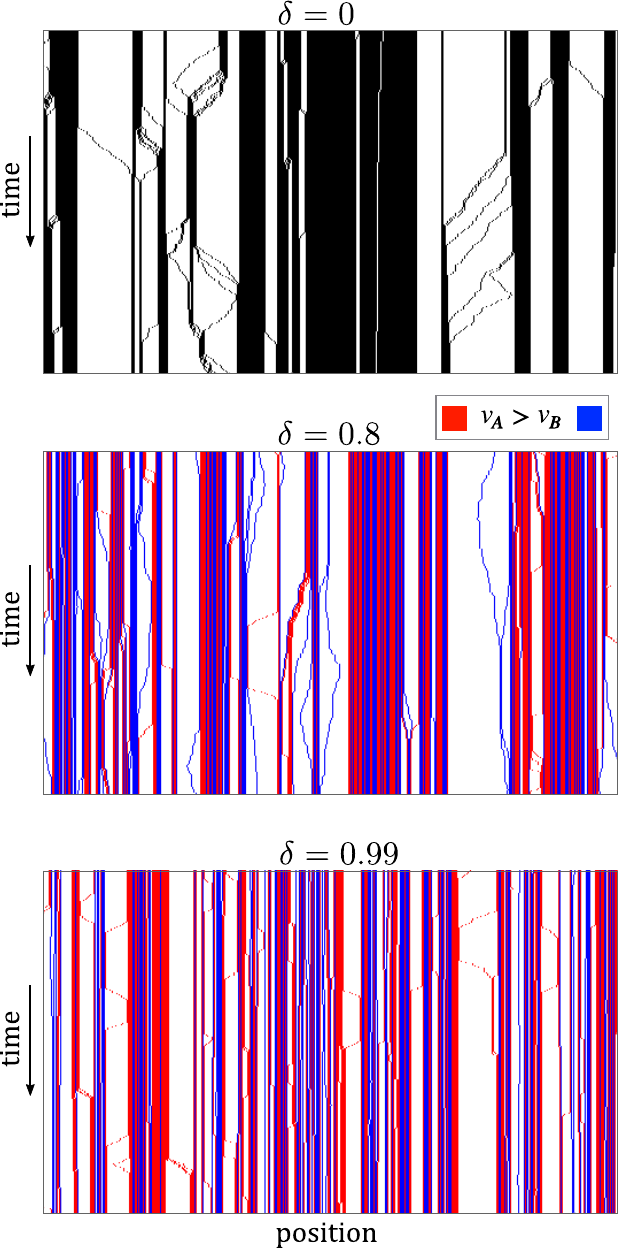}
	\caption{As Fig.~\ref{Fig1}a but for the steady state of a binary system of run-and-tumble particles on a 1D lattice for different bidispersities $\delta$. Only $500$ of $N=2000$ simulated sites are shown. The monodisperse case ($\delta=0$) is shown for comparison. In the bidisperse cases ($\delta>0$), particles with higher (lower) swim speed are in red (blue). The global average speed, tumbling rate, concentration, and composition of the entire simulated system are $\langle v\rangle=v_0=1$, $\alpha=0.01$, $\phi=0.5$, and $50$-$50\%$, respectively.}
	\label{Fig2}
\end{figure}

We start by analysing the binary mixture model for $50$-$50\%$ global composition with $p_{\textrm{s}}=0$, i.e.\ no particle swapping. Fig.~\ref{Fig2} shows snapshots of a section of the 1D simulated system at successive times within the stationary state for different degrees of bidispersity $\delta$. From a homogeneous initial state, particles start to trap each other and form clusters, reaching a steady state characterized by the CSD. For $\delta=0$, all particles are identical and thus have the same tendency to form clusters. In this case, the average cluster size is given by $L_\text{c}\approx \sqrt{2v\phi/[\alpha(1-\phi)]}$ and the CSD is proportional to $\exp{(-l/L_\text{c})}$.\cite{soto2014run} For $\delta>0$, there are two groups of particles, each with a different tendency to form clusters. Distinct typical slopes in the space-time plot indicate distinct swim speeds. Since particle swapping is not yet allowed, the random sequence of particle types does not change in time. The probability to have many particles of the same type at successive positions is vanishingly small, except for small clusters, which are dominated by slower particles as they are more likely to tumble before being absorbed by a cluster.

\begin{figure}[!h]
	\centering
	\includegraphics[width=0.95\columnwidth]{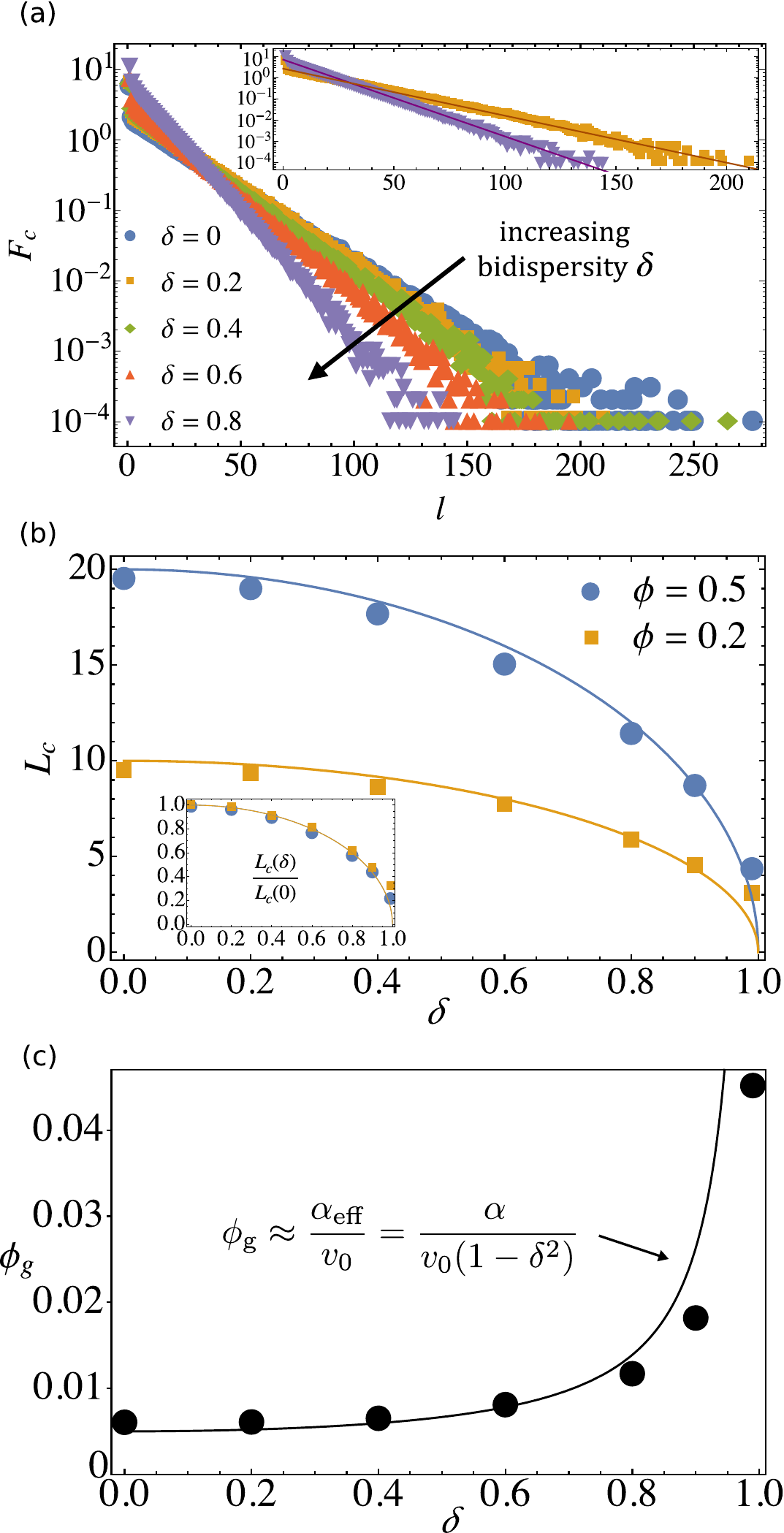}
	\caption{
		(a) Cluster size distribution (log scale) from simulations for various bidispersities $\delta$, with $\langle v\rangle=1$, $\alpha=0.005$, $\phi=0.5$, $N=2000$, and $p_\text{s}=0$. Global composition: $50$-$50\%$. The inset shows the cases $\delta=0.2$ and $\delta=0.8$ (with the same plot markers as in the main figure) compared against the corresponding theoretical results [eqn~\eqref{generalspeedresult}].
		(b) Average cluster size $L_\text{c}$ versus bidispersity $\delta$ at fixed average swim speed, with other parameters as in (a). The points show the simulation results and the lines are the theoretical predictions from Section \ref{theory}. The inset shows the ratio to the corresponding monodisperse case, which is independent of $\phi$. (c) Gas concentration $\phi_\text{g}$ vs.\ $\delta$ with parameters as in (a). Points are the simulation results and the solid line is the theory as obtained by identifying an effective tumbling rate for the mixture.}	
	\label{Fig3}
\end{figure}

The CSD is defined as the average number of clusters of size $l$ and denoted by $F_\text{c}(l)$. Fig.~\ref{Fig3}a shows that as $\delta$ increases the CSD moves towards smaller clusters, even though the global average speed $\langle v\rangle = v_0$ is fixed. The exponential shape of the CSD remains preserved (which allows us to map onto a monodisperse system; see Section~\ref{theory}). Fig.~\ref{Fig3}b shows that $L_\text{c}$ indeed decreases with $\delta$. The inset of Fig.~\ref{Fig3}b has the ratio between the bidisperse and monodisperse values of $L_\text{c}$. This ratio quantifies the clustering reduction by speed diversity. It does not depend on $\phi$ (for $\delta\approx1$, a small dependence on $\phi$ is found, indicating that additional simulation statistics would be required). Similarly, it does not depend on $\alpha$ (data not shown).

Notice that clusters of size $l=1$ correspond to isolated particles and therefore should, in principle, be considered as part of the ``gas'', i.e., not clusters. However, our theory in Section~\ref{theory} \cite{soto2014run} relies on integrating quantities across all positive $l$. Thus, $l=1$ is included in calculating $L_\text{c}$ for a more appropriate comparison. In any case, since at low tumbling rates the gas density is typically small, the contribution to $L_\text{c}$ from isolated particles is likewise very small, as confirmed numerically. The gas concentration $\phi_\text{g}$, which is defined as the average particle concentration in the regions containing only isolated particles (i.e., the regions between clusters of size $\geq2$), increases with $\delta$ (see Fig.~\ref{Fig3}c) since at higher speed diversity fewer particles participate in clusters of size $l>1$, as seen in Fig.~\ref{Fig3}a.

\subsection{Binary mixture with particle swapping}
\begin{figure}[!h]
	\centering
	\includegraphics[width=0.95\columnwidth]{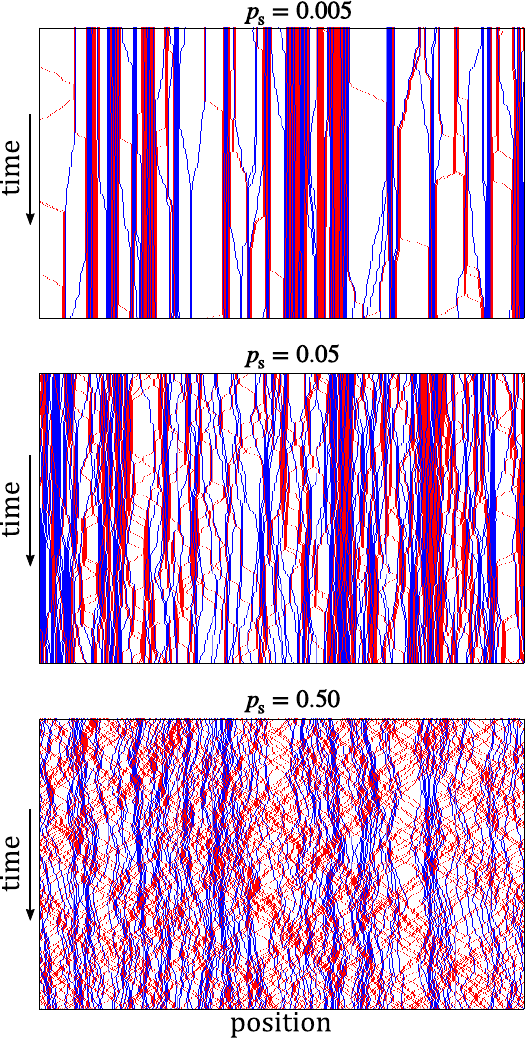}
	\caption{As Fig.~\ref{Fig2} but for fixed $\delta=0.9$ and swapping rate $p_{\rm s}>0$ as indicated, with both particle head-to-head crossing and particle overtaking mechanisms turned on.}
	\label{Fig4}
\end{figure}

With $p_{\textrm{s}}\neq0$, particles have a higher chance to escape from clusters as now this can occur by either tumbling or swapping, where the latter includes both head-to-head crossing and overtaking. Thus, the tumbling rate is effectively increased (see Section \ref{theory}) and therefore cluster formation is further reduced. Fig.~\ref{Fig4} shows snapshots for different $p_{\textrm{s}}$ values. The higher the swapping rate, the more the cluster sizes fluctuate in time.  At high $p_{\textrm{s}}$, the clusters have been mostly destroyed.

\begin{figure}[!h]
	\centering
	\includegraphics[width=0.95\columnwidth]{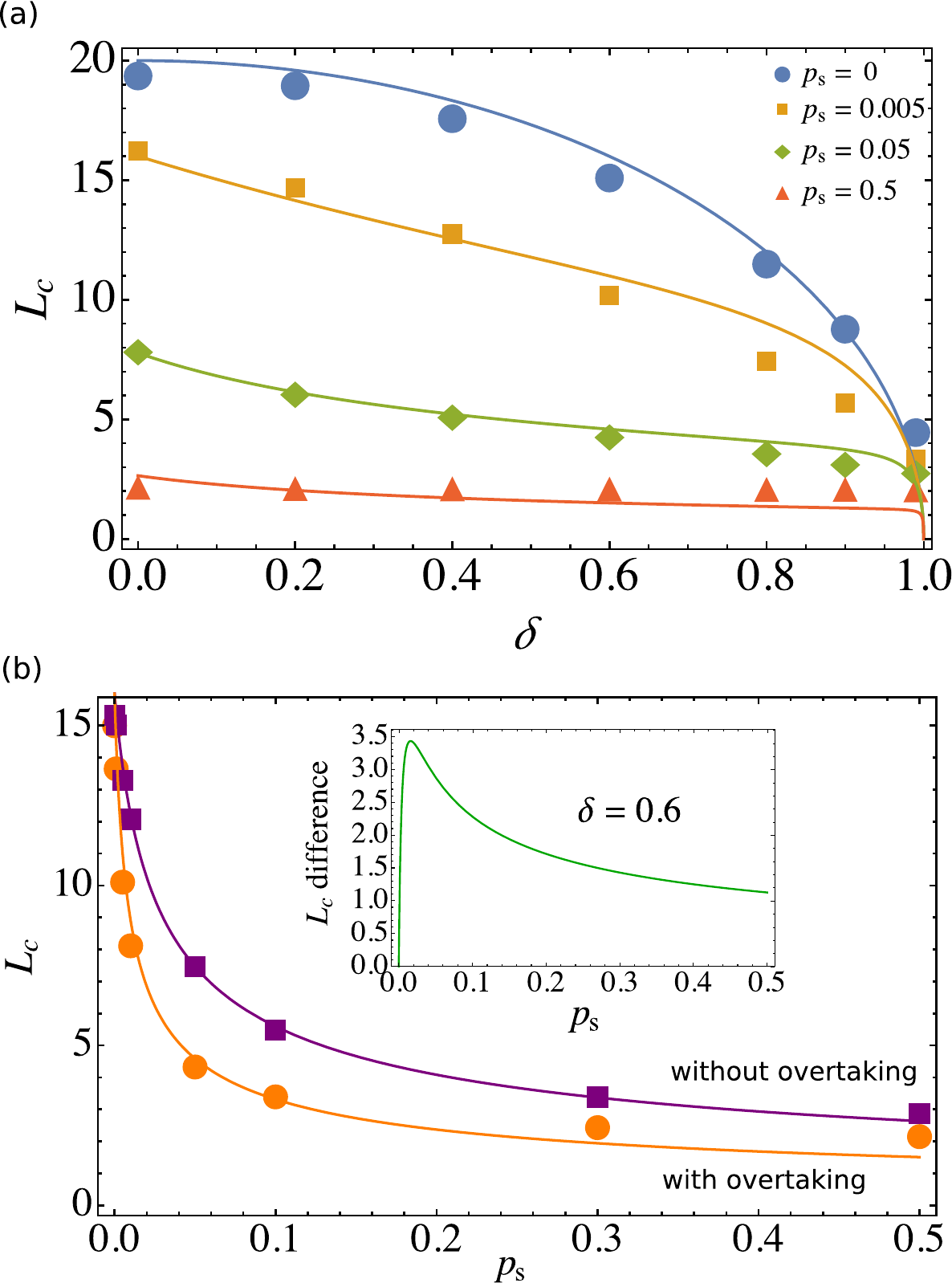}
	\caption{(a) Average cluster size $L_\text{c}$ versus bidispersity $\delta$ at fixed average speed for various particle swapping rates $p_{\rm s}$, from simulations (points) and theoretical predictions for the CSD length scale $L_\text{c}$, eqn.\eqref{generalspeedresult} (lines). (b) Same as in (a) but as a function of $p_{\rm s}\in\{0, 0.001, 0.005, 0.01, 0.05, 0.1, 0.3, 0.5\}$ and comparing the cases with and without particle overtaking. The inset shows the difference between the swapping scenarios. Other parameters as in Fig.~\ref{Fig3}a.}
	\label{Fig5}
\end{figure}

With increasing $p_{\textrm{s}}$ the CSDs recede towards low cluster sizes while approximately maintaining a purely exponential functional form. Fig.~\ref{Fig5}a shows $L_\text{c}$ versus $\delta$ for various values of $p_{\textrm{s}}$, whereas Fig.~\ref{Fig5}b shows $L_\text{c}$ versus $p_{\textrm{s}}$ for fixed values of $\delta$. For high $p_{\textrm{s}}$, the cluster size dependence on $\delta$ is negligible. In this scenario, the particles end up leaving the cluster sooner by swapping than by tumbling. Since tumbling is no longer important, the diversity in the ratio of speed to tumbling rate becomes irrelevant. As a result, the clustering process becomes controlled by swapping events. Fig.~\ref{Fig5}b shows that if overtaking is turned off but head-to-head crossing is kept on, one obtains higher values of $L_\text{c}$, but the overall behaviour with $p_{\textrm{s}}$ is similar. The inset of Fig.~\ref{Fig5}b shows that the difference between these two swapping scenarios peaks at small values of $p_{\textrm{s}}$. This is because at high $p_{\textrm{s}}$ bidispersity becomes irrelevant and therefore overtaking should not be important since its rate is zero without diversity.

\subsection{Fully polydisperse mixture}
\begin{figure}[!h]
	\centering
	\includegraphics[width=0.95\columnwidth]{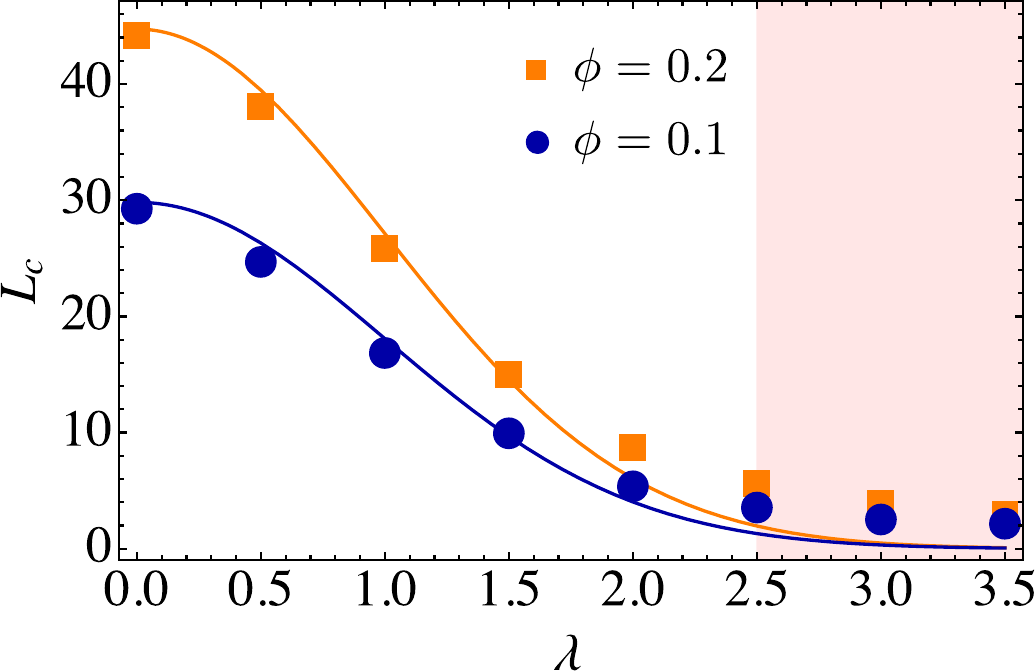}
	\caption{Average cluster size $L_\text{c}$ versus the polydispersity parameter $\lambda$ at fixed average speed $\langle v\rangle=1$ for $\phi=0.1$ and $\phi=0.2$. The points show the simulation results and the lines are the theoretical predictions for the CSD length scale $L_\text{c}$ as given in eqn~\eqref{generalspeedresult}. For the simulations with $\lambda>2.5$, the numerical results no longer agree with the theory, as discussed in Section~\ref{theory}. Other parameters: $\alpha=2.5\times10^{-4}$, $v_\textrm{max}=4$, and $N=10^4$.}
	\label{Fig6}
\end{figure}

For the fully polydisperse distribution of swim speeds in eqn~\eqref{polydist}, we keep $p_{\textrm{s}}=0$ as particle swapping effects are analogous to those in the binary mixture. Fig.~\ref{Fig6} shows the average cluster size. The higher the $\lambda$, the smaller are the clusters, at fixed $\langle v\rangle$. The CSD maintains an approximately purely exponential form becoming almost horizontal at sufficiently high $\lambda$ (data not shown). Additional details for this case are discussed in Section~\ref{theory}.

\section{Theory}
\label{theory}
This section is dedicated to deriving theoretical expressions for the average cluster size $L_\text{c}$ which includes the effect of speed diversity and particle swapping, to be compared to our numerical results presented in Section~\ref{simulations}. We start by reviewing the developments of Ref.~\citenum{soto2014run} (providing previously omitted details) for a lattice model with identical RT particles. We explicitly consider an arbitrary self-propulsion speed $v$ instead of the case $v=1$ of the original derivation. By doing so, we will be able to extend the results to speed-diverse systems. We conclude the section by incorporating particle swapping into the theory, considering both head-to-head crossing and overtaking.  
\subsection{Monodisperse systems}
\label{monotheory}
Following Ref.~\citenum{soto2014run}, we assume that the positions of the borders of a cluster, as well as the cluster itself, are independent random walkers, and that interactions between clusters are weak. Within this approximation (which corresponds to $\alpha/v\ll \phi$ and is used throughout the present work), the authors of Ref.~\citenum{soto2014run} have calculated the distributions of sizes for the clusters and gas regions, $F_{\text{c}}(l)$ and $F_{\text{g}}(l)$, respectively, in the one-component case. (For the purpose of calculating the distribution $F_{\text{g}}(l)$, the gas can be regarded as consisting of empty regions since the concentration of isolated particles is proportional to $\alpha$; see below.) Such distributions were obtained by considering that the clustering process approaches a thermodynamic-like equilibrium where $F_{\text{c}}(l)$ maximizes an entropy for the number of possible cluster configurations. The resulting distributions follow an exponential decay with the region size $l$:  
\begin{equation}
F_{\text{c}}(l) = A_{\text{c}} e^{-l/L_{\text{c}}} ~~ {\rm{and}} ~~ F_{\text{g}}(l) = A_{\text{g}} e^{-l/L_{\text{g}}}.
\label{eq:fcg}
\end{equation}
$A_{\text{c,g}}$ and $L_{\text{c,g}}$ are the parameters of the distribution to be found, with $L_{\text{c,g}}$ corresponding to the stationary average cluster or gas region size. To find these parameters, we first invoke particle conservation, which implies that gas and cluster regions together cover the whole system, i.e., 
\begin{equation}
\sum_l l F_\text{c}(l)+\sum_l l F_\text{g}(l)=N.
\label{eq:ss1}
\end{equation}
Then, since we consider periodic boundary conditions, the number of gas regions must be equal to the number of cluster regions, which leads to 
\begin{equation}
\sum_l F_\text{c}(l)=\sum_l F_\text{g}(l).
\label{eq:ss2}
\end{equation}
Moreover, since the the concentration of isolated particles is $\phi_\text{g}= \alpha/v$ as shown in Ref.~\citenum{soto2014run}, we assume that for high persistence, i.e., high $v/\alpha$, one has $\phi_\text{g}\approx0$. On the other hand, the particle concentration inside the cluster is $\phi_\text{c}=1$ by definition. Taking one typical cluster and one typical gas region, the total amount of particles reads
\begin{equation}
L_\text{c}\phi_\text{c}+L_\text{g}\phi_\text{g}=\left(L_\text{c}+L_\text{g}\right)\phi,
\label{eq:ss3}
\end{equation}
where, due to particle conservation, on the right hand side the size of the cluster and gas region taken together appears multiplied by the system's overall particle concentration $\phi$.

We recall the assumption that cluster-cluster interactions are weak; in fact, these interactions are assumed to occur only through uncorrelated emissions of particles into the gas, and absorptions from there. This implies low tumbling rates (${\alpha/v\ll\phi}$), consistently with the previous steps. In this regime, there are essentially no particles in each gas region ($\phi_\text{g}\approx0$), meaning that these regions are almost empty. To close our system of equilibrium equations, we assume detailed balance (since the clustering process can be mapped onto a thermodynamic-like equilibrium process) and calculate the balance between production and evaporation of dimers (i.e., clusters of size $l=2$). Their formation is controlled by emissions of particles at each boundary of a gas region. In order to compute the production rate of dimers, denoted by $W^+_2$, suppose that a particle at the right boundary of a gas region of typical size $L_g$ is emitted from a cluster with velocity $v$. Such an emission occurs at rate $\alpha/2$. Because tumbling rates are small, we assume that the emitted particle will keep its velocity direction and reach the next cluster after a travel time $L_g/v$. For a new dimer to be created, a particle from the left boundary of the gas region must be emitted, also with probability $\alpha/2$ at each time step, before the first particle's arrival. As for the first particle, we assume that the second emitted particle will move ballistically, i.e., without tumbling, until it encounters the first particle. That is, if the second particle is emitted before the arrival of the first particle, then the dimer will form. For a particle to be emitted at a time $\tau$, it must \textit{not} be emitted before that, i.e., until a time $\tau - 1$. This happens with probability $(1 - \alpha/2)^{\tau - 1} (\alpha/2)$. Therefore, the probability for a new dimer to form is given by 
\begin{align}
\nonumber P_+ &= \left(\frac{\alpha}{2}\right) \left[ \left(\frac{\alpha}{2}\right) \sum_{\tau = 0}^{L_{\text{g}}/v} (1 - \alpha/2)^{\tau - 1}\right], \\
&\approx \left(\frac{\alpha}{2}\right)  \left[1 - (1 - \alpha/2)^{L_{\text{g}}/v}  \right],
\end{align}
where the $\alpha/2$ outside the square brackets accounts for the emission of the first particle, whilst the terms inside the square brackets ensures that the second particle leaves its cluster before the travel time $L_{\text{g}}/v$. Note that, in order to approximate the expression, we used our assumption that $\alpha$ is small. The global production rate of dimers is then given by $W^+_2 = 2 P_+  N_{\text{g}}$, where $N_{\text{g}} = \sum_l F_\text{g}(l)$ is the total number of gas regions and the factor two accounts for the spatially inverted case, where the particle on the left is emitted first. Putting this together yields
\begin{align}
\nonumber W^+_2 &= \alpha \left[ 1 - (1 - \alpha/2)^{L_{\text{g}}/v}\right] N_{\text{g}} \\
& \approx \frac{\alpha^2}{2} \frac{L_{\text{g}}}{v} N_{\text{g}}.
\end{align}

On the other hand, the rate of evaporation of dimers, $W^-_2$, can be calculated as follows. First, note that the only dimers that need to be considered are the ones composed of two particles facing each other, as otherwise the dimer would either have evaporated or be quickly absorbed by other clusters (and so it would not be noticed as a dimer in the steady-state statistics). That is, a dimer is destroyed whenever one of its constituents tumbles to a different direction, which occurs with probability $P_- = \alpha/2$ for each of them, leading to a total probability of evaporation equal to $2P_-$. Then, the global rate of dimer evaporation is obtained by multiplying this total tumbling probability of the dimer by the total number of dimers $F_{\text{c}}(l = 2)$, i.e., $W^-_2 = \alpha A_{\text{c}} e^{-2/L_{\text{c}}} \approx \alpha A_{\text{c}}$, where we used that $2/L_{\text{c}}\approx 0$.
Therefore, the final equilibrium condition used to close our system of equations, $W^-_2 = W^+_2$, amounts to 
\begin{equation}
A_{\text{c}} \approx \frac{\alpha}{2} \frac{L_{\text{g}}}{v} N_{\text{g}}.
\label{eq:ss4}
\end{equation}

To solve the system of equations~\eqref{eq:ss1}-\eqref{eq:ss4} analytically, we replace the summations by integrals from $l=0$ to $\infty$. The solution is
\begin{equation}
A_\text{c} \approx \frac{N\alpha\left(1-\phi\right)}{2v} ~~{\rm and}~~ L_\text{c} \approx l_\text{c}(v,\alpha,\phi)\equiv \sqrt{\frac{2v\phi}{\alpha(1-\phi)}},
\label{monoparam}
\end{equation}
where, due to the small $\alpha$ limit (i.e., $\phi_\text{g}\approx0$), all instances of $v \phi - \alpha$ and $v - \alpha$ have been replaced by $v \phi$  and $v$, respectively.

\subsection{Speed-diverse systems}
To include speed diversity, let us consider a multi-component system with an arbitrary number of particle types $Q$, which are characterized by speeds $v_i$  and are present at global concentrations $\phi_i$ with $i=1,\dots,Q$ and $\sum_i \phi_i=\phi$.  In this case, the size distributions of clusters and gas regions are still given by exponentials. This fact can be verified either via simulations, as shown in Section~\ref{simulations}, or by noticing that the original derivation of the CSD in Ref.~\citenum{soto2014run} becomes specific to the case of identical particles only \textit{after} the exponential shape is obtained. We thus have $F_\text{c}(l) = A_\text{c} e^{-l/L_\text{c}}$ with new constants $A_\text{c}$ and $L_\text{c}$ to determine in the speed diverse case. 
The only expression that requires alteration is the one that explicitly involves self-propulsion speed, that is, the rate of dimer creation. It can now be written as
\begin{equation}
W^+_2\approx\sum_{ij} \alpha\left[1-(1-\alpha/2)^{ L_\text{g}/v_j}\right]N^{ij}_\text{g},
\end{equation}
where the summation runs over particle types and $N^{ij}_\text{g}$ is the \textit{total} number of gas regions bounded by particles of types $i$ and $j$ on the left and right borders, respectively. (Since $N^{ij}_\text{g}$ is symmetric in $i$ and $j$, the term $(1-\alpha/2)^{ L_\text{g}/v_j}$ does not need to be symmetrized.) A cluster border will be occupied by a particle of type $i$ with a probability that depends only on the global concentration $\phi_i$ and $\alpha$, independently of $v_i$. As a result, the chance to find a gas region simultaneously bounded by types $i$ and $j$ obeys $N^{ij}_\text{g}\sim \phi_i \phi_j$, which is then normalized so that $\sum_{ij}N^{ij}_\text{g}$ gives the total number of gas regions. On the other hand, $W^-_2$ remains unchanged since it is speed-independent. The resulting CSD parameters can be expressed, using the function $l_\text{c}$ defined in~\eqref{monoparam}, as
\begin{equation}
A_\text{c} \approx \frac{N\alpha\left(1-\phi\right)}{2v_{\rm eff}}~~\text{and}~~L_\text{c} \approx l_\text{c}(v_{\rm eff},\alpha,\phi),~~\text{with}~~v_{\rm eff}=\left(\sum_i \frac{\phi_i}{\phi v_i}\right)^{-1}
\label{generalspeedresult}
\end{equation}
a \textit{monodisperse} effective speed which is found to be the component-weighted \textit{harmonic} mean speed of the mixture. As such, $v_{\rm eff}$ encodes speed diversity entirely and is the effective self-propulsion speed in the monodisperse case that gives the same CSD as in the speed-diverse case.

We briefly remind the reader that if a vehicle travels a certain distance at speed $v_1$ and returns the \textit{same} distance at speed $v_2$, then its average speed is the harmonic mean of $v_1$ and $v_2$, not the arithmetic mean. The total travel time is the same as if it had travelled the whole distance at that average speed. However, if the vehicle travels for a certain amount of \textit{time} at speed $v_1$  and then the same amount of time at a speed $v_2$, then its average speed is the arithmetic mean of $v_1$ and $v_2$. Therefore, our result can be understood as follows. The average cluster size is ultimately set by the time that an arbitrary particle takes to cross a typical gas region. This time is an arithmetic average over the times taken by each particle type, where we consider that the chance of having a particle of type $i$ travelling in the gas is just proportional to $\phi_i$. Thus, the average speed at which such typical fixed distance is covered is the harmonic average of the speeds, as given in eqn~\eqref{generalspeedresult}, not the arithmetic one. 

For the $50$-$50\%$ binary mixture, our derivation leads to
\begin{equation}
L_\text{c}^{\text{bi}} \approx \sqrt{\frac{2v_0(1-\delta^2)\phi}{\alpha(1-\phi)}},
\label{binaryandfully}
\end{equation}
and, for the fully polydisperse mixtures, the result is
\begin{equation}
L_\text{c}^{\text{poly}} \approx \sqrt{\frac{2v_0 \phi}{\alpha(1-\phi)}}e^{-\frac{\lambda^2}{4}}=\sqrt{\frac{2\langle v\rangle \phi}{\alpha(1-\phi)}}e^{-\frac{\lambda^2}{2}},
\label{binaryandfully2}
\end{equation}
where we have recalled that in the fully polydisperse case $\langle v\rangle =v_0\exp{(\lambda^2/2)}$. Expressions \eqref{binaryandfully} and \eqref{binaryandfully2} are in excellent agreement with the simulation results as presented in Section \ref{simulations}. In particular, the theory predicts that the ratio between the speed-diverse and monodisperse values of $L_\text{c}$ does not depend on $\phi$ or $\alpha$ as shown in the inset of Fig.~\ref{Fig3}b. 

The gas concentration, albeit small and taken to be zero in the analytical derivation of Section \ref{monotheory}, can be written using the monodisperse effective speed $v_\text{eff}$ and extending the monodisperse expression\cite{soto2014run} to obtain ${\phi_\text{g}=\alpha /v_\text{eff}}$ or, equivalently, ${\phi_\text{g}=\alpha_{\rm eff}/\langle v\rangle}$, where $\alpha_{\rm eff}\equiv\alpha \langle v\rangle /v_\text{eff}$ is the effective tumbling rate in the monodisperse case that gives the same CSD as in the speed-diverse case.
The resulting expression is highly accurate, as shown in Fig.~\ref{Fig3}c. Our theory in \eqref{generalspeedresult} becomes less accurate at high speed diversity as a large number of very slow particles arise, in which case the assumption ${\alpha/v\ll\phi}$ is no longer valid.

\subsection{Swapping}

Particle swapping is taken into account by adding appropriate terms to the monodisperse effective tumbling rate $\alpha_\text{eff}$.
This is because particle swapping enables an additional mechanism for cluster-border evaporation, other than tumbling. For simplicity, we consider only the $50$-$50\%$ binary mixture as other cases are analogous. A satisfactory approximation for head-to-head crossing effects is to consider $\alpha_{\rm eff}\to\alpha_{\rm eff}+\kappa_{\rm HC} p_s$, where $\kappa_{\rm HC}$ is a parameter proportional to the fraction of pairs susceptible to head-to-head crossing. In principle, it depends on $p_s$, too. But, to first order in $p_{\rm s}$, it can be considered a constant. Because head-to-head crossing rates are proportional to the relative speed between the particle types, there should be a corresponding dependence on $\delta$, but in the $50$-$50\%$ binary mixture this cancels out because the average of $v_i+v_j$ across particle types is independent of $\delta$.
By fitting data without overtaking for several values of $\delta$ and $p_s$, we find $\kappa_{\rm HC}\approx0.56$.

Proceeding similarly for the case with both head-to-head crossing and overtaking mechanisms, we use 
\begin{equation}
\alpha_{\rm eff}\to\alpha_{\rm eff} +\kappa_{\rm HC}p_s + \kappa_{\rm T} p_s \delta,
\end{equation}
where $\kappa_{\rm T}$ is taken as a constant and the new term is linear in $\delta$ since overtaking depends on the relative speed only between the particles of faster-behind-slower pairs. Fitting data from simulations with overtaking, we obtain $\kappa_{\rm T}\approx1.97$. As shown in Fig.~\ref{Fig5}, this approach provides good results.

\section{Conclusions and discussion}
\label{conc}

In this work we showed how motility-induced self-clustering in confined active matter can be reduced by diversity of self-propulsion speeds. Also, cluster sizes are further reduced by confinement relaxation. We used a minimal quasi-1D discrete lattice model of run-and-tumble particles with a distribution of self-propulsion speeds. Neighbouring particles were allowed to perform head-to-head crossing and overtaking at rates proportional to their relative speeds, depending on whether they are face-to-face oriented or there is a faster particle behind a slower one. This mimics a narrow channel whose width is large enough to allow for some position swapping events. A binary mixture and a fully polydisperse system were studied. Without swapping, the average cluster size $L_\text{c}$ \textit{decreases} with diversity. This is equivalent to a system of identical particles whose speed is the \textit{harmonic} mean speed of the mixture. 
With swapping, particles can escape from clusters more quickly. Consequently, $L_\text{c}$ decreases with swapping rates and depends less strongly on diversity. At sufficiently high swapping rates, clustering then becomes controlled by head-to-head crossing and overtaking events and thus speed diversity becomes irrelevant.
We derived an accurate dynamical equilibrium theory for the CSDs and gas concentrations that is applicable to all models studied here.

In order to calculate $L_c$ for the mixture, at first glance one could be tempted to insert the arithmetic mean self-propulsion speed of the mixture into one-component theories or even to take the arithmetic mean of the one-component $L_c$'s. Here we find that both ideas provide significantly wrong results (more below). Unlike the case of tumbling-rate diversity, we highlight that the above results for speed diversity imply that the average cluster size follows a non-arithmetic generalized power mean of the one-component $L_\text{c}$'s. In fact, our average cluster size result in eqn~\eqref{generalspeedresult} can be rewritten as
\begin{equation}
L_\text{c} \approx \left(\sum_i \frac{\phi_i}{\phi}L_i^p\right)^{1/p}~~\text{with}~~L_i=\sqrt{\frac{2v_i \phi}{\alpha (1-\phi)}}
\end{equation}
and the exponent $p=-2$. This is different from the case of tumbling-rate diversity previously studied where we had $p=1$, that is, the arithmetic mean.\cite{de2021active}  In fact, although for speed diversity an arithmetic mean of the one-component $L_\text{c}$'s would qualitatively capture a decrease in the mixture's $L_\text{c}$, it is quantitatively wrong. Also, notice that the arithmetic mean result for tumbling-rate diversity can in principle also be obtained by the analytical method described in Section \ref{theory}, but in that case we no longer have $N^{ij}_\text{g}\sim \phi_i \phi_j$ as the number of gas regions bounded by a certain particle type depends non-trivially on that particle's tumbling rate (still, we have inserted numerical results for $N^{ij}_\text{g}$ and obtained accurate values for $L_\text{c}$; data not shown).

For the parameters investigated, our data do not show strong ``fractionation'', i.e., clusters with compositions that differ from the global composition. For small clusters, however, we do observe a systematic deviation: small clusters are typically richer in slower particles (data not shown). This is because faster particles travel too fast between clusters to allow for the emission of another particle that could meet it to form a dimer (and then other small cluster sizes). Instead, the first emitted particle is quickly reabsorbed by another cluster before a second particle is emitted. Investigating exactly which parameters would generate strong fractionation in 1D is beyond the scope of our work. This question would be more relevant in 2D, where one might speculate, for instance, that the faster particles would dominate the interior of big clusters since they would be more likely to occupy newly available vacancies.

To estimate whether our qualitative clustering behaviour is also valid for other models of active particles that are not on-lattice or run-and-tumble, one can look at the answer to this question for monodisperse systems. For off-lattice models, simulations show that monodisperse systems of run-and-tumble bacteria, active Brownian particles, and active Ornstein-Uhlenbeck particles, have an average cluster size given by $L_\text{c}\sim\sqrt{\rho u/\omega}$, where $\rho$ is the off-lattice model global concentration, $u$ is the active speed, and $\omega$ is an inverse persistence time parameter.\cite{D0SM00687D} This is precisely the off-lattice version of the length scale expression \eqref{monoparam}. Consequently, our qualitative results for mixtures are indeed expected to be universal across the most used models in 1D scalar active matter.

The clustering reduction effects presented here can indeed be relevant in biological experimental situations. Consider the unicellular microorganisms known as \textit{choanoflagellates}. Although individually they are better described as smooth, active Brownian swimmers, their clustering behaviour is expected to be equivalent to that of run-and-tumble particles because of the equivalence between these models with respect to collective phenomena.\cite{cates2013active} These cells naturally differentiate into slower and faster particles with a speed-diversity standard deviation that can be obtained from literature experimental data as in Ref.~\citenum{sparacino2020solitary}. One can then calculate the error in estimating their $L_\text{c}$ via one-component models that use the arithmetic mean speed instead of the harmonic mean speed, leading to $\Delta L_\text{c}\equiv\left[l_\text{c}\left(v_\text{eff},\alpha,\phi\right)-l_\text{c}\left(\langle v\rangle,\alpha,\phi\right)\right]/l_\text{c}\left(v_\text{eff},\alpha,\phi\right)\approx30\%$.


In future work, it would be relevant to include features such as propulsion mechanisms that are so strong that the particles collectively move and merge clusters. In this kind of scenario, one expects that, for less soft active particles, swapping could become less frequent in narrow channels; at the same time, pushing effects due to direct contact would be enhanced. Also, one could consider lattice models with rules to mimic hydrodynamic interactions, which can substantially change cluster escape times \cite{alarcon2017morphology}. Another interesting future avenue is the case of direction-dependent speeds arising due to external forces.

\section*{Author contributions}
PdC contributed with conceptualization, simulations, theoretical analyses and the writing of the original draft. SD and FMR also helped with the conceptualization and developed most of the analytical derivations. All authors contributed equally to interpreting the simulation data, developing the main ideas of the theory and revising and editing the manuscript.

\section*{Conflicts of interest}
There are no conflicts to declare.

\section*{Acknowledgements}
This research is supported by Fondecyt Grant No.~1180791 (RS) and ANID -- Millennium Science Initiative Program -- NCN19\_170D (RS and PdC), Chile.



\balance

\renewcommand\refname{References}

\bibliography{Active.bib} 
\bibliographystyle{rsc} 

\end{document}